\documentclass[
    ,final            % use final for the camera ready runs
  ]
  {aipproc}

\layoutstyle{6x9}

\begin{document}

\title{Gamma-Rays from Large Scale Structure Formation
and the Warm-Hot Intergalactic Medium: Cosmic Baryometry with Gamma-Rays}

\author{Susumu Inoue}{
  address={Max-Planck-Institut f\"ur Kernphysik, Postfach 103980, 69029 Heidelberg, Germany},
  altaddress={Max-Planck-Institut f\"ur Astrophysik, Karl-Schwarzschild-Str. 1, 85758 Garching, Germany}
}
\author{Masahiro Nagashima}{
  address={Dept. of Physics, Kyoto University, Sakyo-ku, Kyoto 606-8502, Japan},
  altaddress={Dept. of Physics, University of Durham, South Road, Durham DH1 3LE, United Kingdom}
}
\begin{abstract}
It is shown that inverse Compton gamma-rays from electrons accelerated
in large scale structure formation shocks can be crucially affected by
non-gravitational effects such as radiative cooling and galaxy formation,
with corresponding uncertainties by an order of magnitude
in either the gamma-ray source counts or the extragalactic background contribution.
However, this also implies that such gamma-rays may in the near future provide us
with valuable information about the fraction of cosmic baryons in different forms,
particularly the warm-hot intergalactic medium
where the majority of the baryons in the universe are believed to reside.
We address this problem in a simple way through
semi-analytic modeling of structure formation shocks
which self-consistently treats merger and accretion shocks.
\end{abstract}

\maketitle

\section{Introduction}
The majority of the baryons in the universe today are believed to
reside in a warm-hot intergalactic medium (WHIM) at temperatures $T \sim 10^5-10^7$ K,
as a result of shock heating during the hierarchical buildup of
large scale structure in the universe \cite{co99,dav01}.
They are often referred to as `missing baryons',
since quantitative measurements of the WHIM through direct observations are still lacking,
hampered by heavy Galactic obscuration in the relevant extreme UV to soft X-ray bands
(notwithstanding important but fragmentary information from absorption lines
that probe only selected lines of sight \cite{nic04}).
Current indirect estimates of the cosmic fraction of baryons in the WHIM $f_{WH}$
range from $\sim$20 to $\sim$70 \%
\cite{fuk03,fp04,ssp04}.
In view of the significance of elucidating this fundamental component of the universe,
dedicated satellite missions such as
the Diffuse Intergalactic Oxygen Surveyor \cite{oha04}
and the Missing Baryon Explorer \cite{fan03} are being planned
in order to detect emission lines from the WHIM and directly constrain $f_{WH}$.

On the other hand, the same large scale structure formation (SF) shocks
that create the WHIM may give rise to GeV-TeV gamma-ray emission
through nonthermal electron acceleration and inverse Compton upscattering
of the cosmic microwave background.
Such gamma-rays may be observable either as
a contribution to the cosmic gamma-ray background (CGB) \cite{lw00}
or as individual sources \cite{tk00}.
This interesting possibility has spawned a number of studies
using numerical simulations \cite{kes03,min02}
or semi-analytic methods \cite{gb03b,gb04},
although most such works had limited their scope
to treating purely gravitational effects.
In reality, the global hydrodynamical evolution of intergalactic gas
and the associated gamma-ray emission
can be crucially affected by non-gravitational effects such as 
radiative cooling (with consequent star formation and feedback)
and photoionization heating.

By considering such non-gravitational effects in a simplified way,
we show here that there should be a nontrivial connection between
SF gamma-rays and the baryonic fraction in the WHIM.
The problem is addressed through semi-analytic modeling of SF shocks
based on Monte Carlo merger trees with multiple mergers \cite{sk99},
which allows a self-consistent treatment of major and minor merger shocks
as well as diffuse accretion shocks.
The full details can be found in a forthcoming paper (Inoue and Nagashima, in prep.).

\section{Formulation}
An important point in considering nonthermal effects due to SF shocks
is that such shocks can be either strong or weak,
with Mach numbers $\cal{M}$ ranging from
very large ones ($\gg 1$) for minor mergers between systems with large mass ratios
or accretion of relatively cold gas onto a large object,
to values as low as $\sim$1.5--3 in the case of major mergers
between virialized objects of comparable masses \cite{bd03,rkhj04,tak99}.
This implies that the spectral index of shock accelerated particles $p$
can be either the strong shock limit of $p \simeq 2$ or much steeper with $p>2$,
leading to considerably differenct effects at high energies \cite{gb03a}.
Thus it is imperative to account for the distribution of shock Mach numbers appropriately.

One way to address this problem is through
full-scale cosmological hydrodynamical simulations \cite{kes03,min02,rkhj04}.
Here we opt for a semi-analytic approach
based on the extended Press-Schechter (PS) formalism of structure formation \cite{lc93},
which gives a simple yet reasonably accurate description of
the hierarchical gravitational growth of dark matter halos.
In particular, we employ the multiple merger tree algorithm
of Somerville \& Kolatt \cite{sk99},
which accurately reproduces the total and conditional halo mass functions
and also accounts for diffuse accretion.
At each time step in the merger tree, we also employ an accurate mass function
derived from very high resolution N-body simulation results \cite{yny04}.
Note that the semi-analytic model of Gabici \& Blasi (GB03) \cite{gb03a,gb03b,gb04}
is built on the simpler binary merger tree algorithm of Lacey \& Cole \cite{lc93},
which cannot treat accretion and
is known to produce self-inconsistent results
when the merger tree is extended to high redshifts \cite{sk99}.
(Nevertheless, it is found that the differences are not very large at low redshifts,
and our results below for SF gamma-rays are in basic agreement with \cite{gb03b,gb04}
when appropriate comparisons are made.)

Our basic assumptions are as follows.
(1) The adopted cosmological parameters are
$\Omega_m$=0.3, $\Omega_\Lambda$=0.7, $\Omega_b$=0.044 and $h$=0.7.
The normalization and spectral index of primordial fluctuations are respectively
$\sigma_8$=0.9 and $n$=1.
(2) At each step, a multiple merger event between more than two halos
is pictured as an ensemble of binary mergers
with the primary (i.e. most massive) progenitor.
Associated with each binary pair are two shocks propagating within them.
Mass below a certain mass scale (see below) in the timestep
accretes spherically onto the primary.
(3) An effective Mach number is assigned to each merger shock.
The temperature of the preshock gas is the virial value for the relevant progenitor.
The relative infall velocity $v$ is given by
$v^2=2G (M_{A+B}) [(f_0 (M_B/M_A)+1)/(R_A+R_B)-1/2R_{AB}]$,
where $M_i$ and $R_i$ are the masses and virial radii for
A, B and A+B denoting the two progenitors and merged halo, respectively.
This is similar to \cite{gb03a},
except that we include a parameter $f_0$
that is calibrated to match the simulation results for major mergers \cite{tak99}.
Diffuse accretion shocks are always strong.
(4) The electron injection efficiency is fiducially a fraction $\xi_e=0.05$
of the energy dissipated at the shock, i.e. the difference between
the post-shock and pre-shock thermal energies.
The injection spectral index $p$ is related to the shock Mach number $\cal{M}$
as $p=2({\cal{M}}^{2}+1)/({\cal{M}}^{2}-1)$ (test particle assumption).
The emitted inverse Compton spectrum is a broken power-law,
with energy indices $(p-1)/2$ and $p/2$ respectively above and below the cooling break energy
where the electron cooling time equals the shock crossing time,
and a high energy cutoff where the cooling time equals the acceleration time.
Only primary electrons are considered.
(5) The gas fraction inside halos is $\Omega_b/\Omega_m$ when affected solely by gravity.
Non-gravitational effects due to radiative cooling and photoionization heating
which can be effective in certain mass ranges
are incorporated in a simplified way as described below.

Two characteristic mass scales are important concerning non-gravitational effects.
One is the scale of the post-reionization Jeans mass (more accurately the filtering mass),
below which gas cannot appreciably cool and collapse inside virialized halos
due to photoionization heating by the UV background \cite{gne00}.
This may be considered the natural boundary
distinguishing diffuse accretion and clumpy merging
and is identified with the mass resolution scale in our merger tree algorithm.
Taking its velocity dispersion $V_{acc}$ as a parameter,
a realistic range may be $V_{acc} \simeq 20-70$ km/s 
as suggested by detailed modeling \cite{gne00}.
We assume a fiducial value of $V_{acc}=40$ km/s,
corresponding to mass $3.0 \times 10^{10} M_\odot$ at $z=0$.
The other important scale is the maximum cooling mass
above which gas cannot significantly cool because of
the reduced cooling function in the pertinent temperature range;
its velocity dispersion is parameterized by $V_{cut}$.
In halos between $V_{acc}$ and $V_{cut}$,
gas can cool efficiently and condense into stars, i.e. become galaxies.
The observed galaxy luminosity function indicates $V_{cut} \simeq 150-250$ km/s,
and we fiducially take $V_{cut}=200$ km/s
or $3.7 \times 10^{12} M_\odot$ at $z=0$ in terms of mass.
SF shocks and associated emission will be suppressed
in systems which have converted a large fraction of its gas into stars,
and we treat this effect simply by removing SF shocks in a fraction $f_{GF}$ of halos 
between $V_{acc}$ and $V_{cut}$ and account only for mass growth through their merging.

An interesting connection can be made between
our principal parameters $V_{acc}$, $V_{cut}$, $f_{GF}$,
and the present-day fraction of baryons in the universe in different forms.
Following \cite{dav01} in dividing cosmic baryons into four phases,
diffuse ($T<10^5$ K), condensed (stars and cold gas),
warm-hot ($10^5<T<10^{7}$ K), and hot ($T>10^{7}$ K),
these respectively relate in our picture to systems with velocity dispersion $V<V_{acc}$,
a fraction $f_{GF}$ of $V_{acc}<V<V_{cut}$,
the rest $1-f_{GF}$ of $V_{acc}<V<V_{cut}$ plus a part of $V>V_{cut}$ with $T<10^7$ K,
and the remainder of $V>V_{cut}$ with $T>10^7$ K.
If we take our fiducial values $V_{acc}=40$ km/s and $V_{cut}$=200 km/s,
there is a one to one relation between $f_{GF}$ and $f_{cond}$,
the baryon fraction condensed into stars and cold gas.
This relation can be quantitatively evaluated
using the semi-analytic galaxy formation model of \cite{ny04}.
In turn, $f_{cond}$ can be related to $f_{WH}$
by subtracting the baryon fractions in the diffuse and hot phases.
For example, cosmological simulations
including radiative cooling and galaxy formation \cite{dav01}
indicate a range $f_{cond} \simeq 0.2-0.4$ and $f_{WH} \simeq 0.2-0.4$ at $z=0$,
which corresponds to $f_{GF} \simeq 0.6-0.9$.
Alternatively, a recent observational census \cite{fuk03}
suggests $f_{cond} \simeq 0.1$ and $f_{WH} \simeq 0.4$, 
which is consistent with $f_{GF} \simeq 0.4$.

A further non-gravitational effect
that might be important is feedback (pre-)heating by
supernovae-driven winds or AGN outflows,
as indicated by the observed X-ray scaling relations of groups and clusters
(\cite{sp03} and references therein).
Since the details of such processes are highly uncertain at present,
we defer a consideration of these effects to future work
(see \cite{ti02} for an early, crude discussion).

One important caveat is in order concerning our formulation.
In the PS picture, all the dark matter in the universe is described
as being bound inside spherically virialized halos of some mass.
This is a fairly good approximation, as many comparisons with N-body simulations demonstrate
(e.g. \cite{yny04} and references therein.)
However, the same cannot be said about the gas component.
In fact, we have explicitly assumed the fraction of gas with $V < V_{acc}$
to be in diffuse form outside bound halos due to photoionization heating.
It is less clear how much of the WHIM, particularly the gas with $V_{acc} < V < V_{cut}$,
can be considered to be inside or outside bound objects.
Although hydrodynamical simulations indicate that a large part of the WHIM arises
through shock heating by gravitational infall
onto filamentary or sheet-like structures \cite{co99,dav01},
much of the gas in such structures may actually be interpreted as residing inside
sufficiently small halos if seen at high enough resolution.
Since the essential driving force of WHIM evolution is the gravity of the dark matter,
most of which is indeed in bound form, we have chosen to describe the WHIM
as gas inside bound haloes with the corresponding range of virial temperatures
which do not condense into stars.
Just how good such a description (or some alternative, e.g. \cite{fl04}) may be
can only be judged through future comparisons with detailed numerical simulations.

\section{Results and Discussion}

\begin{figure}
  \includegraphics[height=.5\textheight]{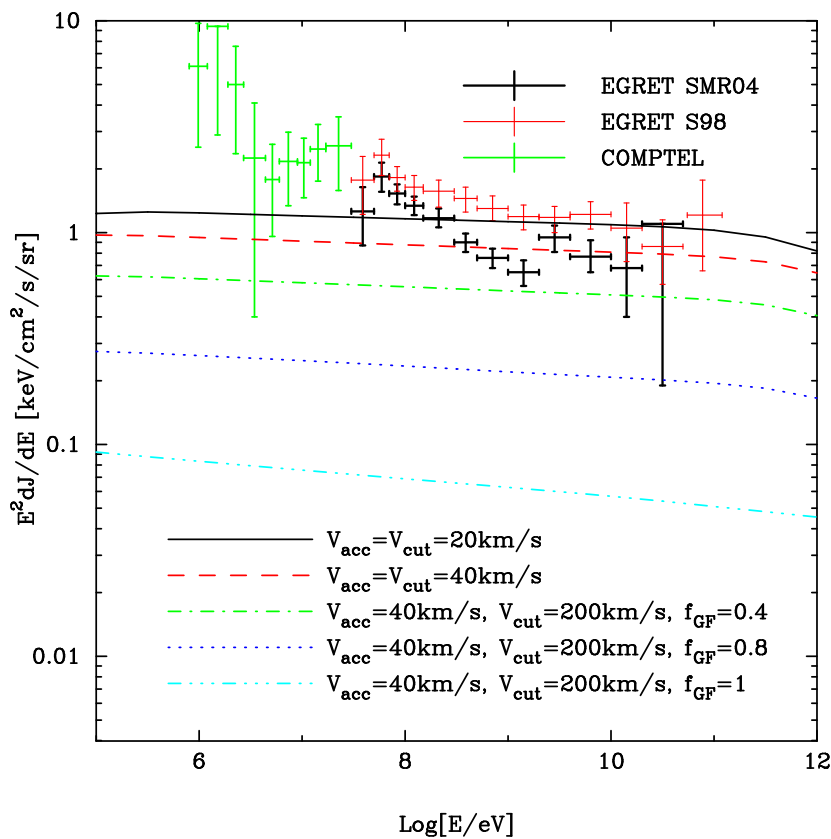}
  \caption{
Cosmic gamma-ray background for different values
of $V_{acc}$, $V_{cut}$ and $f_{GF}$ as indicated in the legend,
compared with COMPTEL and EGRET (both S98 and SMR04) data.
}
\end{figure}

Figure 1 shows our results of the SF shock contribution to the CGB
for different values of $V_{acc}$, $V_{cut}$ and $f_{GF}$.
To be compared are CGB data from COMPTEL \cite{wei00}
and EGRET, including both the old Sreekumar et al. (1998; S98) \cite{sre98} 
and new Strong, Moskalenko \& Reimer (2004; SMR04) \cite{smr04} determinations.

We first take $V_{acc}=V_{cut}=20$ km/s, corresponding closest
to the situations treated in numerical simulations,
where the sole non-gravitational effect is a temperature floor of $T=10^4$ K
\cite{kes03,min02}.
The result accounts for almost all of the S98 CGB, and in fact exceeds the new SMR04 CGB,
indicating either that this case does not represent reality
or that $\xi_e$ is significantly less than the fiducial value 0.05.
Although this is in more agreement with the result of \cite{min02} than of \cite{kes03},
here we reserve a quantitative judgement,
given the approximate nature of our formulation.

For this and all other cases discussed here,
the end result is dominated by minor merger shocks,
with accretion shocks amounting to at most 1\% of the S98 CGB.
Keeping $V_{acc}=V_{cut}$ (i.e. no condensation into stars),
a larger value decreases the merger component and hence the total CGB,
while slightly increasing the accretion component; for smaller values, vice-versa.
Taking $f_{GF}=1$ (complete condensation in the cooling regime)
with $V_{acc}=40$ km/s fixed, varying $V_{cut}$ has a dramatic effect,
with the CGB being suppressed by more than an order of magnitude
as $V_{cut}=200$ km/s is approached.
This can be understood as removing larger and larger galaxy-scale objects,
which can potentially produce strong shocks in minor mergers with cluster-scale objects.
Our fiducial set of $V_{acc}=40$ km/s, $V_{cut}=200$ km/s leads to
$\sim 10$ \% of the S98 CGB, consistent with the results of GB03.

Obviously, when condensation occurs only for
a fraction $f_{GF}$ of objects in the cooling regime,
the reduction is less, and one gets a CGB somewhere between 10 to 100 \% of the S98 CGB.
Recalling the above-mentioned connection between $f_{GF}$ and $f_{cond}$,
the current uncertainty in $f_{cond} \simeq 0.1-0.4$ allows a range $f_{GF}=0.4-0.9$,
and the CGB due to SF shocks cannot be reliably predicted to within an order of magnitude.
However, this points to an interesting possibility
of constraining $f_{cond}$ and hence $f_{WH}$
if the SF shock contribution to the CGB can be observationally determined.
In practice, this requires removing other contributions (e.g. blazars) to the CGB
with good precision, which may not be an easy task.

\begin{figure}
  \includegraphics[height=.5\textheight]{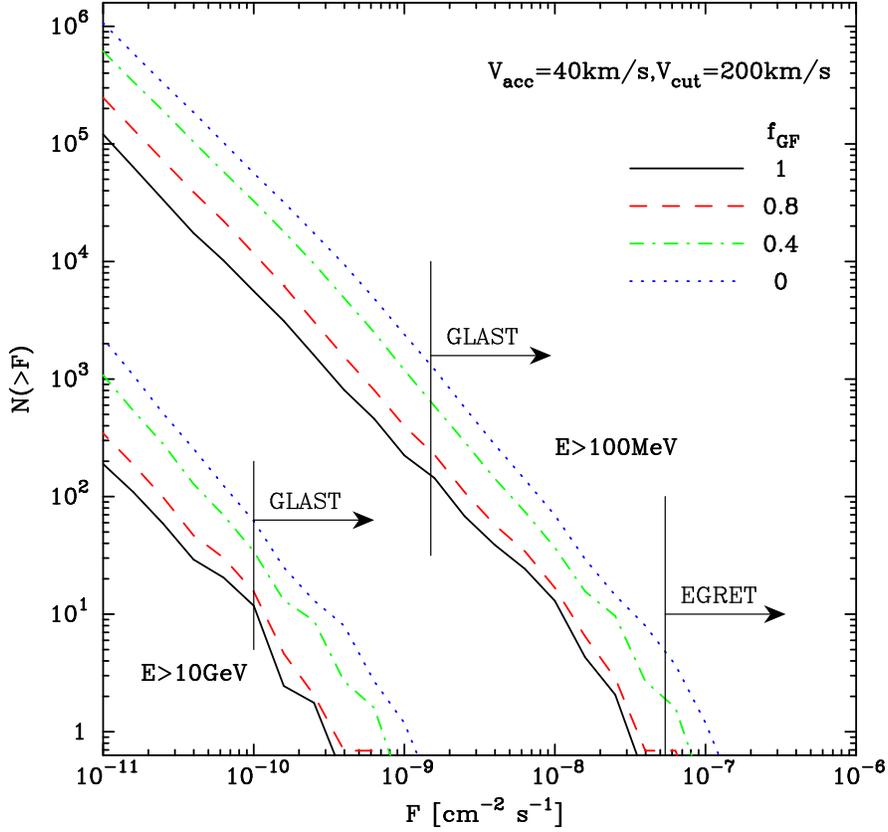}
  \caption{
Gamma-ray source counts at 100 MeV and 10 GeV for different values of $f_{GF}$,
compared with the sensitivities of EGRET and GLAST.
}
\end{figure}

A more promising way to constrain $f_{WH}$ with gamma-rays
may be through the statistics of source counts.
Figure 2 displays the cumulative source counts due to SF shocks
at energies > 100 MeV and > 10 GeV, for fixed fiducial values of $V_{acc}$ and $V_{cut}$
and different $f_{GF}$.
Again, differences of an order of magnitude can be seen,
depending on how much minor merger shocks are suppressed.
For $f_{GF}=0.9$ ($f_{cond} \simeq 0.25$, $f_{WH} \simeq 0.25$),
$\sim 100$ and $\sim$ 10 sources should be observable by GLAST
at >100 MeV and >10 GeV, respectively,
while none exists above the EGRET sensitivity at >100 MeV.
For $f_{GF}=0.4$ ($f_{cond} \simeq 0.1$, $f_{WH} \simeq 0.4$),
the respective numbers are $\sim 600$ and $\sim 30$ at >100 MeV and >10 GeV
for GLAST, whereas a few are expected for EGRET.
The fact that EGRET actually saw no emission associated with clusters \cite{rei03}
may point to either $f_{GF}>0.4$ or $\xi_e <0.05$.
This underlies the need for the electron injection effiency to be pinned down,
preferably through detailed observations of individual sources
where the kinetic energy can be estimated independently.
Once this is done,
SF gamma-rays may provide an indirect but very valuable probe
of the unknown fraction of baryons in the WHIM.
In fact, a close connection between SF gamma-rays and the WHIM
is not surprising at all, as they both arise from the same large-scale shocks.
However, the quantitative correspondence is a nontrivial one
involving the reduction of strong, minor merger shocks by condensation into stars.

To summarize, we have investigated
inverse Compton gamma-rays from large scale SF shocks
including non-gravitational effects
with a self-consistent semi-analytic formulation.
Radiative cooling and galaxy formation
were shown to have crucial impact,
with the predicted CGB contribution and gamma-ray source counts
uncertain by an order of magnitude
depending on the fraction of baryons condensing into stars.
This in turn implies that SF gamma-rays may serve
as an indirect `baryometer' of the universe
and probe the `missing' fraction of baryons in the WHIM,
which is very difficult to measure directly.

The present work is an example of nonthermal phenomena due to SF shocks
where semi-analytic, PS-based methods can be applied effectively,
allowing the exploration of physical effects in a simple way
which is not always the case with numerical simulations.
However, in view of the numerous approximations in our formulation,
further studies with simulations are warranted,
both to confirm the qualitative trends found here
and to make predictions that are quantitatively more robust.
The effects of feedback (pre-)heating, which have not been treated here,
may also be potentially important and need to be investigated further.

\end{document}